\documentclass[prl, twocolumn,10pt,longbibliography,superscriptaddress]{revtex4-2}
\usepackage{textcomp}
\usepackage{amsmath}
\usepackage{amsfonts}
\usepackage{calc}
\usepackage{mathrsfs}
\usepackage{amssymb}
\usepackage{amsmath}
\usepackage{array}
\usepackage{color}
\usepackage{bm}
\usepackage{graphicx}
\usepackage{hyperref}
\usepackage{booktabs}
\usepackage[scr=boondoxo,scrscaled=1.05]{mathalfa}
\usepackage[dvipsnames]{xcolor}
\usepackage[normalem]{ulem}

\newcommand{\e}{\epsilon}

\begin{document}
	\title{Gravitational waveforms for compact binaries from second-order self-force theory}
	
\author{Barry Wardell}
\affiliation{School of Mathematics and Statistics, University College Dublin, Belfield, Dublin 4, Ireland, D04 V1W8}
\author{Adam Pound} 
\affiliation{School of Mathematical Sciences and STAG Research Centre, University of Southampton, Southampton, United Kingdom, SO17 1BJ}
\author{Niels Warburton}
\affiliation{School of Mathematics and Statistics, University College Dublin, Belfield, Dublin 4, Ireland, D04 V1W8}
\author{Jeremy Miller}
\affiliation{Department of Physics, Ariel University, Ariel 40700, Israel}
\author{Leanne Durkan}
\affiliation{School of Mathematics and Statistics, University College Dublin, Belfield, Dublin 4, Ireland, D04 V1W8}
\author{Alexandre Le Tiec}
\affiliation{Laboratoire Univers et Th\'{e}ories, Observatoire de Paris, CNRS, Universit\'{e} PSL, Universit\'{e} de Paris, F-92190 Meudon, France}

\date{\today}

\providecommand{\NW}[1]{{\textcolor{Red}{\texttt{NW: #1}}}}
\providecommand{\AP}[1]{{\textcolor{Red}{\texttt{AP: #1}}}}
\providecommand{\LD}[1]{{\textcolor{Red}{\texttt{LD: #1}}}}
\providecommand{\BW}[1]{{\textcolor{Red}{\texttt{BW: #1}}}}
\providecommand{\ALT}[1]{{\textcolor{Red}{\texttt{ALT: #1}}}}
\providecommand{\ALTco}[1]{{\textcolor{Red}{\texttt{ALT comment: #1}}}}
\providecommand{\oops}[1]{{\textcolor{Red}{#1}}}
\providecommand{\done}[1]{{\textcolor{Green}{\texttt{\checkmark~#1}}}}
\providecommand{\notdone}[1]{{\textcolor{Red}{\texttt{\checkmark~#1}}}}

\renewcommand{\e}{\epsilon}

\begin{abstract}
We produce gravitational waveforms for nonspinning compact binaries undergoing a quasicircular inspiral. Our approach is based on a two-timescale expansion of the Einstein equations in second-order self-force theory, which allows first-principles waveform production in tens of milliseconds. Although the approach is designed for extreme mass ratios, our waveforms agree remarkably well with those from full numerical relativity, even for comparable-mass systems. Our results will be invaluable in accurately modelling extreme-mass-ratio inspirals for the LISA mission and intermediate-mass-ratio systems currently being observed by the LIGO-Virgo-KAGRA Collaboration.
\end{abstract}

\maketitle

\textit{Introduction.}
The era of gravitational-wave astronomy is upon us, with the LIGO-Virgo-KAGRA (LVK) Collaboration now routinely observing dozens of gravitational wave signals. In its most recent release, LVK announced the detection of 35 signals from the merger of compact-object binaries seen during the second half of its most recent observing run \cite{LIGOScientific:2021djp}.

The majority of signals observed by LVK to date have been well represented by existing modelling approaches including numerical relativity (NR) \cite{Duez:2018jaf}, post-Newtonian (PN) theory \cite{Blanchet:2013haa}, the effective one-body (EOB) formalism \cite{Ossokine:2020kjp, Nagar:2018zoe,Cotesta:2018fcv,Cotesta:2020qhw}, NR surrogates~\cite{Varma:2018mmi,Varma:2019csw}, and phenomenological models~\cite{Pratten:2020fqn,Pratten:2020ceb,Garcia-Quiros:2020qpx}. However, LVK is now beginning to see glimpses of binary systems that push the limits of those approaches. One signal, GW191219\_163120, thought to have come from the merger of a $\sim 1:26$ mass ratio binary, was outside the region of parameter space where existing models have been validated, and LVK concluded that there may be systematic uncertainties in their results for that signal as a consequence \cite{LIGOScientific:2021djp}.

In parallel to the efforts of the LVK Collaboration, members of the LISA Consortium are preparing for the space-based Laser Interferometer Space Antenna \cite{LISA:2017pwj}, which will be sensitive to low-frequency, millihertz gravitational wave signals outside of the LVK sensitivity band. LISA will observe new categories of sources including supermassive black hole binaries---involving a pair of comparable-mass supermassive black holes---and extreme-mass-ratio inspirals (EMRIs)---consisting of a stellar-mass compact object in orbit around a supermassive black hole \cite{Babak-etal:17}. EMRIs, in particular, are expected to be well outside the range of mass ratios and orbital configurations accessible to existing waveform models used by the LVK Collaboration.

The primary method of modelling EMRIs is the gravitational self-force approach~\cite{Barack:2018yvs,Pound:2021qin}, which solves the Einstein field equations using an expansion in powers of the mass ratio $\epsilon = m_2/m_1\leq1$, where  $m_1$ and $m_2$ are the binary's constituent masses. At zeroth order in $\epsilon$, the companion moves on a geodesic of the primary's spacetime, and at higher orders it experiences a self-force that drives its inspiral. It is widely expected~\cite{LISA:2022kgy,Amaro-Seoane:2022rxf} that this expansion must be carried to second order in $\epsilon$ to achieve the necessary accuracy for LISA. That expectation is motivated by the fact that over an inspiral, the gravitational waveform phase has an expansion of the form~\cite{Hinderer:2008dm,Pound:2021qin} 
\begin{equation}\label{eq:1PA phase expansion}
\phi = \epsilon^{-1}\phi_0 + \epsilon^0\phi_1+{\cal O}(\epsilon).
\end{equation}
The ``adiabatic" (0PA) term $\phi_0$ involves dissipative effects of the first-order self-force (along with geodesic effects), and the ``first post-adiabatic'' (1PA) term $\phi_1$ involves  dissipative effects of the second-order self-force (along with first-order conservative effects). The 1PA contribution to Eq.~\eqref{eq:1PA phase expansion} is nonnegligible for all mass ratios, but the ${\cal O}(\epsilon)$ error term (which would involve third-and-higher-order effects) \emph{is} negligible for sufficiently small $\epsilon$, suggesting that second order is both necessary and sufficient for EMRI modelling. Recent estimates have suggested 1PA models may even suffice to cover much of the \emph{non}-extreme binary parameter space~\cite{vandeMeent:2020xgc,Ramos-Buades:2022lgf}, further motivating the pursuit of such models.

\begin{figure*}[htb!]
	\includegraphics[width=0.95\textwidth]{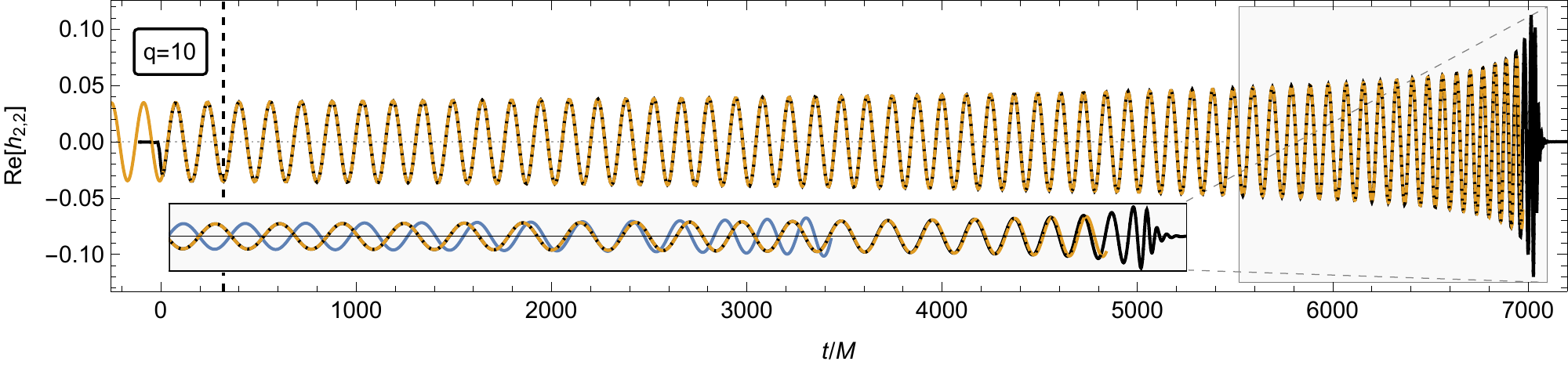}
	\caption{1PAT1 waveform for a mass ratio $1:10$ nonspinning binary (orange). Also included for comparison is the waveform for the same binary produced using an NR simulation (black) \cite{SXS:BBH:1107}. The inset shows a zoomed region near the merger and also shows the corresponding 0PA waveform (blue). The waveforms are aligned in time and phase at $t=320M$, where $1/x \approx 13.83$.}
\label{fig:q10}
\end{figure*}

In this Letter we present the first-ever calculation of these 1PA waveforms by solving the Einstein equations through second order in the mass ratio for nonspinning binaries in quasicircular orbits. Our model also performs remarkably well for more comparable mass ratios -- see Fig.~\ref{fig:q10} -- which broadly validates the predictions of Ref.~\cite{vandeMeent:2020xgc}.

Our model assumes the larger body is a nonspinning black hole, while its companion can be any (nonspinning) compact object. We use geometrized units with $G=c=1$. We also define the large mass ratio $q=1/\epsilon$ and symmetric mass ratio $\nu = m_1 m_2/M^2$, where $M=m_1 + m_2$ is the total mass. Our method begins from expansions in powers of $\epsilon$ at fixed $m_1$, but we consistently re-expand our results in powers of $\nu$ at fixed $M$. This restores in the perturbative solution the inherent discrete symmetry of the full solution under the interchange $m_1 \leftrightarrow m_2$ of the two bodies \cite{LeTiec:14}, and yields the most accurate results in the comparable-mass regime~\cite{Fitchett:1984qn,Anninos:1994gp,Favata:2004wz,Sperhake:2011ik,LeTiec:2011bk,LeTiec-etal:12b,LeTiec:2013uey,Nagar:2013sga,LeTiec:2017ebm,Rifat:2019ltp,vandeMeent:2020xgc}.

\textit{Two-timescale expansion.} We utilize the second-order self-force framework developed in Refs.~\cite{Pound:12a,Pound:12b,Pound-Miller:14,Warburton-Wardell:14,Wardell-Warburton:15,Pound:15c,Miller-Wardell-Pound:16,Pound:17,Miller:2020bft,Spiers_in_prep,Miller_in_prep,Durkan:2022fvm}, building particularly on the two-timescale expansion in Ref.~\cite{Miller:2020bft} and the results reported in~\cite{Pound:2019lzj,Warburton:2021kwk}. A detailed explanation of our model is given in Ref.~\cite{Albertini:2022rfe}, so we restrict ourselves here to a summary of the most pertinent points. The model treats $\epsilon$ as a small parameter, decomposing the spacetime metric as $\mathbf{g}_{\alpha\beta} = g_{\alpha\beta} + h_{\alpha\beta}$, where $g_{\alpha\beta}$ is the Schwarzschild metric of the large black hole and $h_{\alpha \beta}\sim \epsilon$. The perturbation is expanded through order $\epsilon^2$ in terms of slowly evolving amplitudes $h_{\alpha \beta}^{(n,m)}$ and oscillatory phase factors $e^{-im\phi_p}$,
\begin{equation}\label{tt expansion}
  h_{\alpha \beta} = \sum_{m=-\infty}^\infty \big[\epsilon h_{\alpha \beta}^{(1,m)}(J_A,x^i) + \epsilon^2 h_{\alpha \beta}^{(2,m)} (J_A,x^i)  \big]e^{-i m \phi_p},
\end{equation}
where $x^i=(r,\theta,\phi)$ are spatial coordinates.

All time dependence is encoded in the parameters $J_A=(m_1,\Omega)$---the black hole's mass and the companion's orbital frequency $\Omega:=d\phi_p/dt$---and in the orbital phase $\phi_p$. $J_A$ slowly evolves due to gravitational-wave emission and absorption; in this evolution, an $n$PA approximation includes all terms contributing through order $\epsilon^{n+1}$.

The amplitudes $h_{\alpha \beta}^{(n,m)}$ are determined by using separation of variables to decompose the first- and second-order Einstein equations into radial ordinary differential equations for the spherical-harmonic modes of $h_{\alpha \beta}^{(n,m)}$ \cite{Warburton-Wardell:14,Wardell-Warburton:15,Miller:2020bft}. After solving on a grid of $J_A$ values and storing $h_{\alpha \beta}^{(n,m)}$, we can rapidly generate waveforms by solving evolution equations for $J_A$ and $\phi_p$ and evaluating Eq.~\eqref{tt expansion} at future null infinity.

\textit{1PA equations of motion.}
Although the evolution of $m_1$ (and of the black hole's spin, which slowly grows to $\sim\epsilon$) formally appears at 1PA order, the effects are numerically subdominant~\cite{Martel:2003jj} so we choose to neglect them. We thus only need to consider the evolution of the orbital frequency, $\Omega$.
We determine an equation for $d\Omega/dt$ by employing an energy balance law, combining the results of~\cite{Pound:2019lzj,Warburton:2021kwk} for the flux $\mathcal{F}_\infty$ of gravitational-wave energy to infinity and the binding energy $E_{\rm bind}(J_A,m_2)$.
Taking a time derivative of $E_{\rm bind}(J_A,m_2)$ and rearranging, we obtain a set of orbital evolution equations, 
\begin{equation}\label{eq:nonpert}
  \frac{d\Omega}{dt} = -\left(\frac{\partial E_{\rm bind}}{\partial\Omega}\right)^{-1} \mathcal{F}, \quad \frac{d \phi_p}{dt} = \Omega,
\end{equation} 
where ${\cal F}={\cal F}_\infty+\epsilon^2{\cal F}_{\cal H}^{(1)}$, and $\epsilon^2{\cal F}^{(1)}_{\cal H}=dm_1/dt+\mathcal{O}(\epsilon^3)$ is the standard leading-order energy flux into the black hole~\cite{Hughes:1999bq,Fujita:2009us,BHPToolkit}; other than this contribution to the balance law, $m_1$'s evolution is a strictly 1PA effect, and we henceforth treat $m_1$ as constant. A full discussion of Eq.~\eqref{eq:nonpert} and its neglect of numerically small 1PA effects can be found in Sec.~IIB  of Ref.~\cite{Albertini:2022rfe}.

The flux and binding energy are computed from the amplitudes $h^{(n,m)}_{\alpha\beta}$~\cite{Pound:2019lzj,Warburton:2021kwk}. We expand them in powers of $\nu$ at fixed $M$ as $\mathcal{F} = \nu^2\mathcal{F}^{(1)}(x)+\nu^3\mathcal{F}^{(2)}_\infty(x)+\mathcal{O}(\nu^4)$~\cite{Warburton:2021kwk} and $E_{\rm bind} = \nu M \left[\hat E_0(x)+\nu \hat E_{\rm SF}(x)+\mathcal{O}(\nu^2)\right]$~\cite{Pound:2019lzj}, where $x := (M \Omega)^{2/3}$ is a dimensionless measure of the inverse separation, $\mathcal{F}^{(1)}(x) = \mathcal{F}^{(1)}_\infty(x) + \mathcal{F}^{(1)}_{\cal H}(x)$, and $\hat E_0(x) = (1-2x)/\sqrt{1-3x} - 1$. The self-force contribution $\hat{E}_\text{SF}(x)$ to the binding energy is evaluated using the prediction from the first law of compact binary mechanics \cite{LeTiec-etal:12a,LeTiec-etal:12b}, which provides an excellent approximation over the entire inspiral phase \cite{Pound:2019lzj}. Here and below, a numerical superscript $(n)$ indicates a quantity computed from $h^{(n',m)}_{\alpha\beta}$ up to $n'=n$, while a numerical subscript $n$ denotes the PA order at which a quantity contributes.
Fully expanding Eq.~\eqref{eq:nonpert} in powers of $\nu$, we then arrive at our first time-domain post-adiabatic (1PAT1) model for the orbital evolution, 
\begin{align}
  \label{eq:PAT1}
  \frac{d\Omega}{dt} = \frac{\nu}{M^2}\left[ F_0(x)+\nu F_1(x)\right], \quad
  \frac{d \phi_p}{dt} = \Omega,
\end{align}
where $F_0(x)=a(x)\mathcal{F}^{(1)}(x) $ and $F_1(x)=a(x)\mathcal{F}^{(2)}_\infty(x) -a(x)^2\mathcal{F}^{(1)}(x)\partial_{\hat\Omega} \hat E_{\rm SF}$, with $\hat\Omega:=M\Omega$ and $a(x):=-(\partial \hat E_0/\partial\hat\Omega)^{-1}$.

Eq.~\eqref{eq:PAT1} can be rapidly integrated to generate a trajectory, but the 1PAT1 model has the disadvantage of requiring the mass ratio to be chosen \emph{before} solving the equations of motion. An even more efficient approach is to expand the orbital phase and frequency as $\phi_p(\tilde{t},\nu)  = \nu^{-1}[ \phi_0(\tilde{t}) + \nu \phi_1(\tilde{t}) + \mathcal{O}(\nu^2) ]$ and $\Omega(\tilde t,\nu)  = \Omega_0(\tilde t) + \nu \Omega_1(\tilde t) + \mathcal{O}(\nu^2)$, where $\tilde t := \nu t$. Substituting this into Eq.~\eqref{eq:PAT1} and re-expanding, we obtain a second time-domain post-adiabatic model (1PAT2), comprising a hierarchical sequence of equations in which the dependence on the mass ratio has been fully factored out,
\begin{subequations}
\begin{align}
  \label{eq:PAT2}
  \frac{d\Omega_0}{d\tilde t} = \frac{F_0(x_0)}{M^2},\quad \frac{d \phi_0}{d\tilde t} = \Omega_0,
\end{align}
\vspace{-0.5cm}
\begin{equation}
    \frac{d\Omega_1}{d\tilde t} = \frac{1}{M^2}\left[F_1(x_0)+\Omega_1\partial_{\Omega_0} F_0(x_0)\right], \quad 
  \frac{d \phi_1}{d\tilde t} = \Omega_1,
\end{equation}
\end{subequations}
where $x_0:=(M\Omega_0)^{2/3}$. These can be integrated in advance, and their solutions stored, without specifying a mass ratio. Trajectories can then be immediately generated for any given mass ratio.

Although 1PAT2 has an advantage in terms of computational efficiency, comparisons with NR reveal that 1PAT1 has a distinct advantage in terms of accuracy. This is typical of asymptotic methods: expansions at fixed values of determinative dynamical variables ($\Omega$ in this case) are more accurate than expansions at fixed values of extrinsic time parameters ($\tilde t$ in this case).
The 1PAT2 model also breaks down earlier in the inspiral than 1PAT1, as it directly involves the relationship $\Omega_0(\tilde t)$ of an adiabatic model (0PA, given by the equations for $\Omega_0$ and $\phi_0$ alone), which for comparable-mass binaries becomes singular significantly earlier than the 1PAT1 model.

In addition to time-domain waveforms, the two-timescale approach naturally lends itself to the production of frequency-domain waveforms. Writing $\phi_p'(\Omega) = \Omega / \Omega'(t)$, substituting in the 1PAT1 equations for $\Omega'(t)$, and re-expanding in $\nu$, we arrive at a frequency-domain post-adiabatic (1PAF1) model: $\phi_p=\nu^{-1}[\tilde\phi_0(\Omega)+\nu\tilde\phi_1(\Omega)+\mathcal{O}(\nu^2)]$, where
\begin{align}
  \label{eq:PAF1}
  \frac{d\tilde\phi_0}{d\Omega} = \frac{M^2\Omega}{F_0(x)}, \quad
  \frac{d\tilde\phi_1}{d\Omega} = -\frac{M^2\Omega\,F_1(x)}{F_0(x)^2}.
\end{align}
This combines the accuracy advantage of 1PAT1 with the efficiency advantage of 1PAT2. Moreover, frequency-domain waveforms can be especially convenient for data analysis purposes \cite{LIGOScientific:2019hgc}.

\textit{Waveform.}
In order to extract a waveform from our post-adiabatic models we write the strain in terms of a certain tetrad component of our metric perturbation, $h \equiv h_+ - i h_\times = \lim_{r\to \infty} \frac{r}{M} h_{\bar{m}\bar{m}}$ where $\bar{m}^\alpha = \frac{1}{\sqrt{2}r}(0,0,1, - i \csc \theta)$. Decomposing into spin-weight $-2$ spherical harmonic $(\ell,m)$ modes and expanding in the symmetric mass ratio yields
\begin{equation}\label{eq:hlm}
   h_{\ell m} = \Big\{\nu h_{\ell m}^{(1)} + \nu^2 \Big[h_{\ell m}^{(1)} +h_{\ell m}^{(2)}-\frac{2x}{3}\dfrac{dh_{\ell m}^{(1)}}{dx}\Big] \Big\}e^{-i m \phi_p},
\end{equation}
where $h_{\ell m}^{(n)}=h_{\ell m}^{(n)}(x)$ is the asymptotic amplitude from Eq.~\eqref{tt expansion} [i.e., the $(\ell,m)$ mode of $\lim_{r\to\infty}\frac{r}{m_1}h^{(n,m)}_{\bar m\bar m}$], evaluated at $J_A=(M,0,\Omega)$. In 1PAT2, the amplitudes are further expanded as $h_{\ell m}^{(n)}(x)=h_{\ell m}^{(n)}(x_0)+\nu\Omega_1\partial_{\Omega_0}h_{\ell m}^{(n)}(x_0)+\mathcal{O}(\nu^2)$.

\textit{Results.}
In Fig.~\ref{fig:q10} we plot the real part of the 1PAT1 waveform for a mass ratio $q=10$ binary (similar plots for mass ratios $q=1$, $4$, $6$ and $9.2$, as well as a comparison with a post-Newtonian waveform, are given as supplemental material \cite{supplemental-material}). Our waveform (orange) agrees remarkably well with an equivalent waveform produced using an NR simulation (black) until very close to merger, at which point the assumption of adiabaticity breaks down. In contrast, a 0PA waveform (shown in blue in the inset) produced using only the leading-order-in-$\nu$ contributions falls badly out of phase much earlier. 

\begin{figure}[tb!]
	\includegraphics[width=\columnwidth]{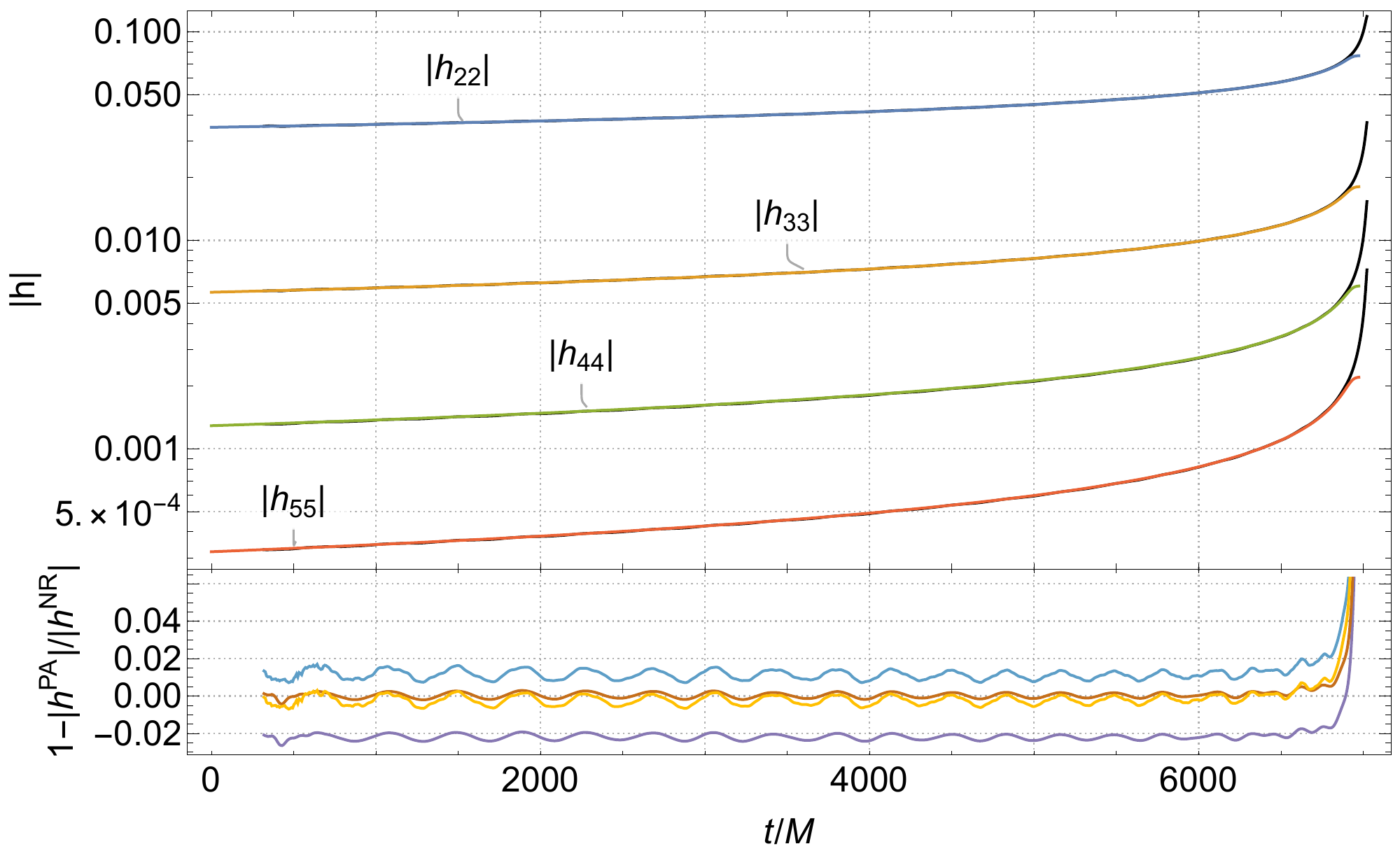}
	\caption{\textit{Top:} Waveform amplitude for a mass ratio 1:10  nonspinning binary for a set of spherical harmonic modes. Also included for comparison is the same waveform produced using an NR simulation \cite{SXS:BBH:1107} (black). \textit{Bottom:} Relative error in the waveform amplitude for the $(\ell,m)=(3,3)$ and $(5,5)$ modes with (red and orange) and without (purple and light blue) re-summation.}
\label{fig:amplitudes}
\end{figure}

\begin{figure}[tb!]
	\includegraphics[width=\columnwidth]{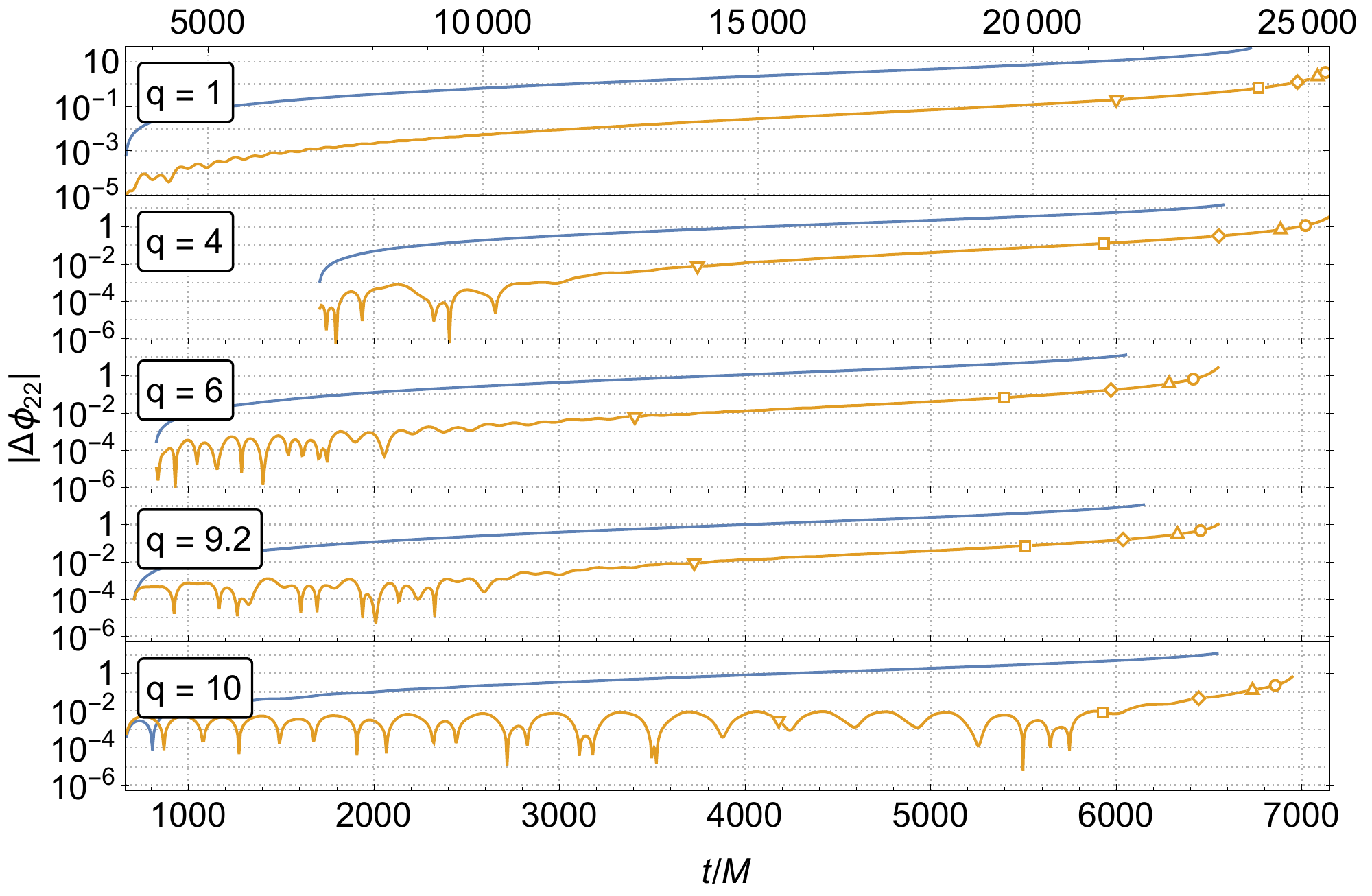}
	\caption{Phase difference between 0PA (blue) / 1PAT1 (orange) and NR waveforms for mass ratios 1:1, 1:4, 1:6, 1:9.2 and 1:10 (in order from top to bottom) binaries for the $(\ell,m)=(2,2)$ mode. The markers indicate the point on the waveform $32$ (upside-down triangles), $16$ (squares), $10$ (diamonds), $6$ (triangles), and $4$ (circles) cycles before merger.}
\label{fig:phase}
\end{figure}

Decomposing the complex waveform modes into a (real) amplitude and phase, we see from Figs.~\ref{fig:amplitudes} and \ref{fig:phase} that both are well captured by our model. This conclusion holds not only for the dominant $(\ell,m)=(2,2)$ mode, but also for higher modes. For the $(2,2)$ mode with $q=10$ we find that the phase error is less than $0.2$ radians and the amplitude error is essentially unmeasurable (i.e. is less than uncertainties in the NR simulation) until $\sim 4$ waveform cycles before merger.


The bottom panel of Fig.~\ref{fig:amplitudes} shows that the odd-$m$ mode amplitudes agree slightly less well with NR results. This can be traced to our perturbative solution's failure to capture the inherent discrete antisymmetry of the odd-$m$ modes under the interchange $m_1 \leftrightarrow m_2$ of the two bodies, as is fully explicit in PN waveforms \cite{Blanchet:2013haa,Kidder:2007rt}. We can restore that antisymmetry by multiplying and dividing by $(m_1-m_2)/M = \sqrt{1-4\nu}$, re-expanding the denominator, and truncating at order $\nu^2$ to obtain amplitudes $\sqrt{1-4\nu}[\nu h^{(1)}_{\ell m} +\nu^2 ( h^{(2)}_{\ell m}+2h^{(1)}_{\ell m})]$. This ``resummation" yields the small but appreciable improvement in accuracy seen in the bottom panel of Fig.~\ref{fig:amplitudes}. (However, we stress that no resummations are used in any other figures.)

\begin{figure}[tb!]
	\includegraphics[width=\columnwidth]{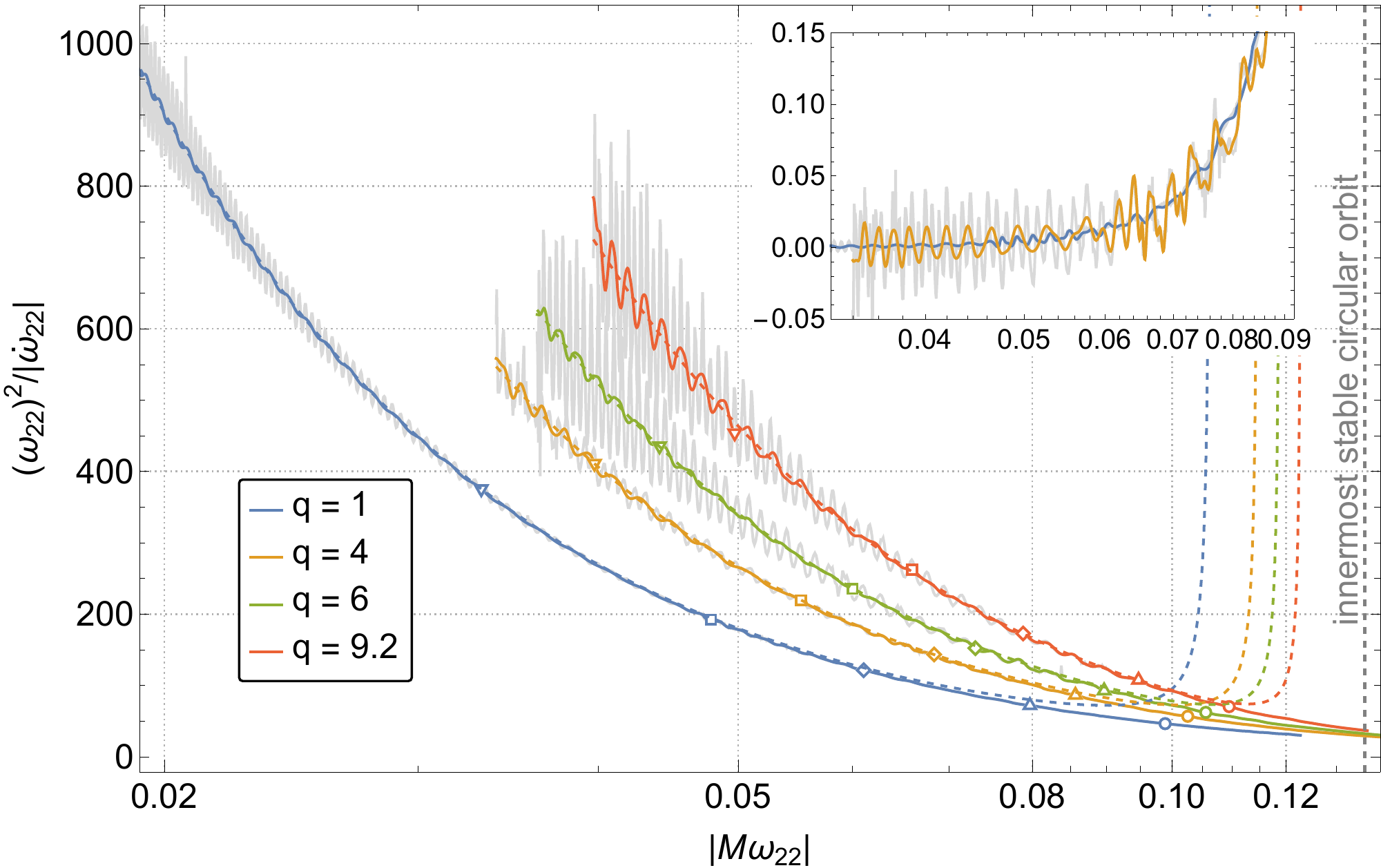}
	\caption{Comparison of $\omega^2/\dot{\omega}$ between NR \cite{SXS:BBH:1132,SXS:BBH:1220,SXS:BBH:0181,SXS:BBH:1108} (solid lines) and 1PAT1 (dashed lines) waveforms for the $(\ell,m)=(2,2)$ mode for a set of mass ratios. The markers are as in Fig.~\ref{fig:phase}. The inset shows the difference in $\dot{\omega}/\omega^2$ between NR and 1PAT1 scaled by $\nu^{-3}$. A low-pass filter was applied to the NR data to eliminate high-frequency numerical noise; versions without the filter applied are shown in light gray.}
	 \label{fig:omega2-omegadot}
\end{figure}

As an alternative to computing the phase difference as a function of time as a measure of accuracy of our waveform, it is useful to compute the dimensionless adiabaticity parameter $\dot{\omega}/\omega^2$ or its inverse~\cite{Baiotti:2010xh}. When this parameter becomes large, the orbit is evolving rapidly and our adiabatic assumption breaks down. The integral of its inverse, $\omega^2/\dot{\omega}$, with respect to $\log \omega$ yields the accumulated phase so this is an indirect measure of the phase accuracy of our waveforms. By plotting the adiabaticity parameter as a function of waveform frequency, $\omega$, we completely eliminate the sensitive dependence of the comparison on any particular choice of point at which to align the phase and frequency of two waveforms. Figure~\ref{fig:omega2-omegadot} shows the inverse adiabaticity parameter computed in NR and the 1PAT1 model as a function of $\omega$ (on a logarithmic scale) for a range of mass ratios. We see that NR and 1PAT1 agree very well until just a few waveform cycles before merger, at which point the adiabatic assumption has broken down and we have no reason to trust our two-timescale models.

\begin{figure}[tb!]
	\includegraphics[width=\columnwidth]{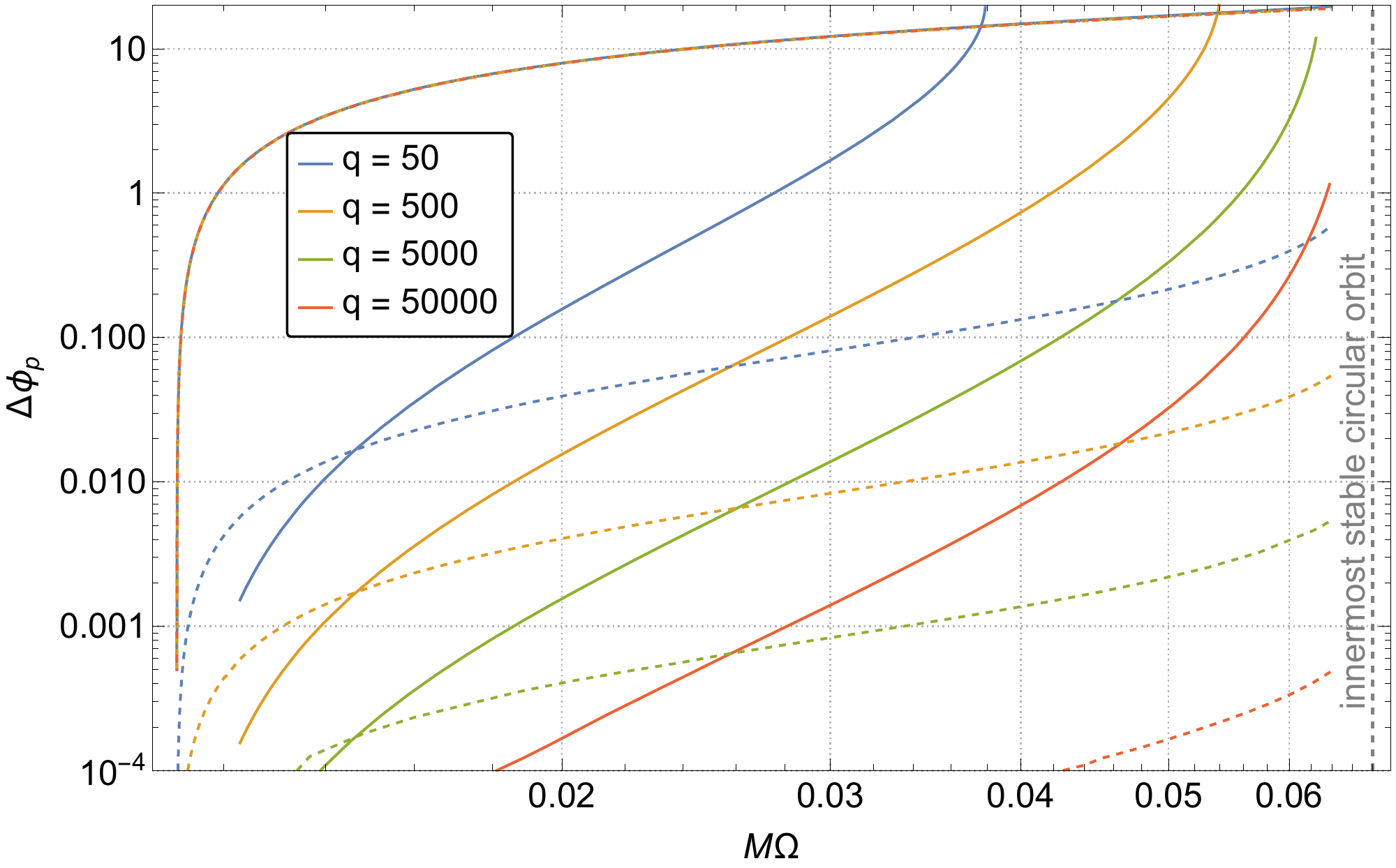}
	\caption{Orbital phase difference as a function of orbital frequency comparing 0PA against 1PAT1 (overlapping dashed curves), 1PAT2 against 1PAT1 (solid curves), and 1PAF1 against 1PAT1 (dotted curves).}
\label{fig:pa-phase}
\end{figure}

We can also internally assess the accuracy of our models by comparing each of our three models for the orbital phase (0PA, 1PAT1, and 1PAT2). Figure \ref{fig:pa-phase} shows the result of this comparison for a range of small mass ratios that are outside the reach of current NR simulations. Although these do not tell us about the absolute error in our models, the comparison does solidify the notion that 0PA waveforms are totally inadequate. So-called ``kludge" models often used for EMRI data analysis~\cite{Chua:2017ujo} are even less accurate, as they represent approximations to the 0PA waveform. Our faster 1PAT2 model agrees modestly well with the 1PAT1 model, with the agreement improving with mass ratio, as expected given that the difference between the models is $\mathcal{O}(\nu)$. However, if one is interested in phase accuracy within a fraction of a radian close to the test-particle innermost stable circular orbit (ISCO), it is clear that the 1PAT2 model will only be useful in that case for $q \gtrsim 50\,000$.

Finally, we highlight the speed of waveform generation within our setup.
The functions on the right-hand side of the 1PA equations of motion are non-oscillatory and, as such, the ordinary differential equations in Eqs.~\eqref{eq:PAT1}-\eqref{eq:PAF1} can be numerically integrated very rapidly, on the order of milliseconds.
The waveform is then constructed by evaluating Eq.~\eqref{eq:hlm} at the relevant sample times and summing over spherical harmonic modes.
It was recently shown that for waveforms containing orders of magnitude more modes than appear in our quasi-circular setup this summation and sampling step can be performed in a 0PA model in few hundred milliseconds for, e.g., year-long EMRI waveforms containing $\sim3.15 \times 10^6$ time steps \cite{Chua:2020stf,Katz:2021yft}.
The structure of the equations at 1PA is the same as at 0PA, so 1PA waveform generation is just as rapid.

\textit{Conclusions.}
The waveforms computed in this work represent the first major result in a programme to produce post-adiabatic waveforms for compact binaries. Although restricted to the case of nonspinning systems and quasicircular orbits, it is nonetheless likely that these waveforms will be useful in their own right. The LVK Collaboration has begun to observe systems with mass ratios well outside the comfort zone of existing waveform models \cite{LIGOScientific:2021djp}, while at the same time being well within the range of validity of our post-adiabatic model.

There are several natural future directions in which to take this work. These primarily revolve around treating more generic configurations, for example where one or both of the compact objects are spinning, or where the orbit may be eccentric and/or precessing.
Another important area of improvement is the treatment of the very last stages of the waveform. Our adiabaticity assumption breaks down as the system approaches the ISCO, where we must transition to a plunge model \cite{Ori-Thorne:00,Apte:2019txp,Burke:2019yek,Compere:2021zfj} followed by a quasinormal mode ringdown \cite{Hatsuda:2020egs}. 

Improvements might also be made to our 1PA inspiral model. We have three sources of uncertainty in our 1PA evolution equations: our use of the first-law binding energy, our neglect of 1PA effects of the primary black hole's evolution, and our neglect of the 1PA horizon absorption ${\cal F}^{(2)}_{\cal H}$. These are analyzed in detail in Ref.~\cite{Albertini:2022rfe}. While their impact is small, it is potentially nonnegligible. In future work we will eliminate them by using the second-order dissipative self-force and the full set of 1PA evolution equations~\cite{Miller:2020bft}.

\begin{acknowledgments}
\textit{Acknowledgements.} We thank Alessandro Nagar and Angelica Albertini for comments on an early draft of this letter. AP acknowledges support from a Royal Society University Research Fellowship, a Royal Society Research Fellows Enhancement Award, and a Royal Society Research Grant for Research Fellows. NW acknowledges support from a Royal Society - Science Foundation Ireland University Research Fellowship via grants UF160093 and RGF\textbackslash R1\textbackslash180022. This work makes use of the Black Hole Perturbation Toolkit \cite{BHPToolkit} and Simulation Tools \cite{SimulationToolsWeb}.
\end{acknowledgments}

\bibliography{bibfile}

\end{document}

\title{Supplemental Material}
	
	\maketitle

\section{Configurations}

In Table~\ref{table:configs} we list each of the configurations studied along with the corresponding SXS simulation. These correspond to the additional mass ratios shown in the phase plot, Fig.~4, and adiabaticity plot, Fig.~5 of the main letter. In each case we have aligned the waveforms in time such that at a matching time $t_{\rm match}$ the orbital frequency is $\Omega_{\rm match}$, the separation is $1/x_{\rm match}$, and the frequency of the $(\ell,m)=(2,2)$ mode of the waveform is $\omega_{\rm match}$. We also introduced a constant phase shift such that the waveform phases agree at the matching time.

Note that there are oscillations in the NR waveform frequency at early times. As a result, the frequency of the NR waveform is not exactly $\omega_{\rm match}$ at the matching time. We have, however, ensured that the frequency evolution of the NR waveform agrees well over an extended period of length several hundread $M$, which is significantly longer than the period of the oscillations in the NR waveform frequency.
\begin{table}[htb!]
  \centering
  \begin{tabular}{llllll}
    \toprule
    Mass ratio & \textbf{$t_{\rm match}$} & $1/x_{\rm match}$ & $\Omega_{\rm match}$ & $\omega_{\rm match}$ & SXS Simulation\\
    \midrule
                 1   & 170 & 23.9515 & 0.0085310 & 0.0170635 & SXS:BBH:1132 \cite{SXS:BBH:1132} \\
                 4   & 340 & 15.9188 & 0.0157447 & 0.0314960 & SXS:BBH:1220 \cite{SXS:BBH:1220} \\
                 6   & 400 & 14.7158 & 0.0177142 & 0.0354358 & SXS:BBH:0181 \cite{SXS:BBH:0181} \\
                 9.2 & 340 & 13.8227 & 0.0194585 & 0.0389239 & SXS:BBH:1108 \cite{SXS:BBH:1108} \\
                 10  & 320 & 13.8279 & 0.0194476 & 0.0389017 & SXS:BBH:1107 \cite{SXS:BBH:1107} \\
    \bottomrule
  \end{tabular}
  \caption{Details of the configurations and matching parameters for each of the 5 mass ratios included in our comparisons.}
  \label{table:configs}
\end{table}

\newpage

\section{Comparisons between NR and 0PA and 1PA GSF waveforms}

In this section we show the comparison between NR and GSF waveforms for mass ratios $q=\{1,4,6,9.2,10\}$. These are similar to Fig.~1 in the main letter, but for different mass ratios. 
\begin{figure}[!htb]
	\includegraphics[width=0.95\textwidth]{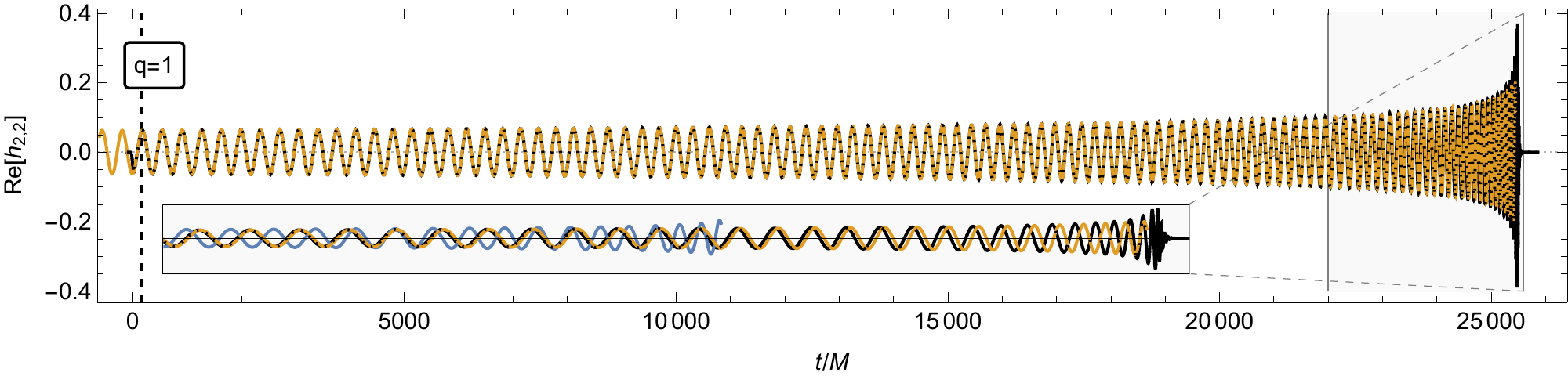}
	\includegraphics[width=0.95\textwidth]{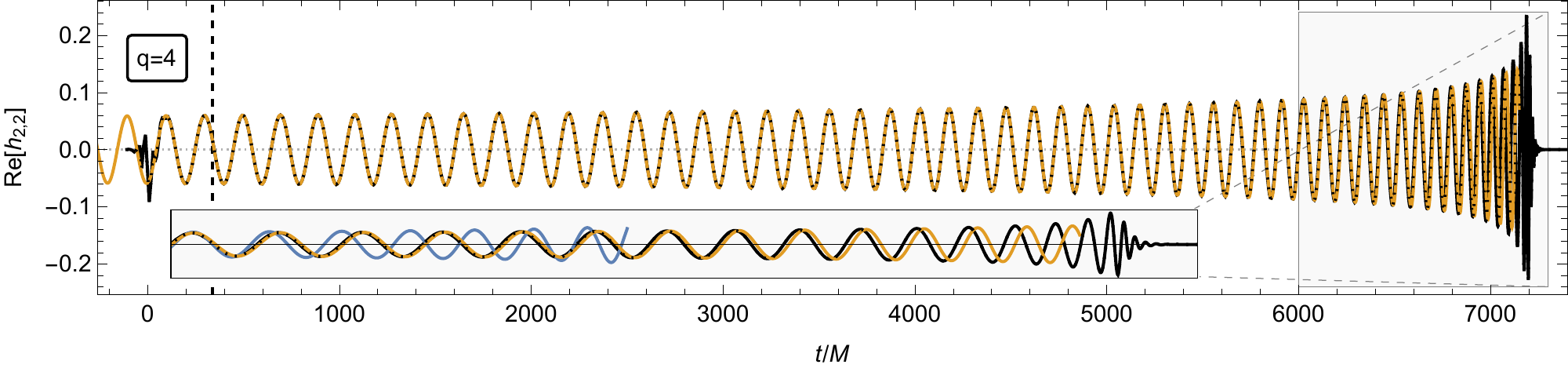}
	\includegraphics[width=0.95\textwidth]{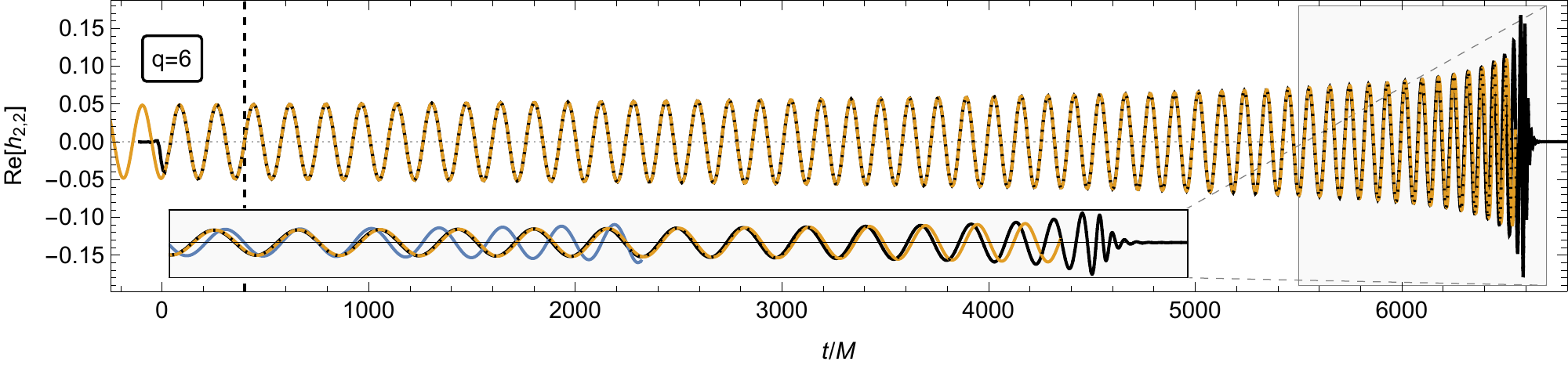}
	\includegraphics[width=0.95\textwidth]{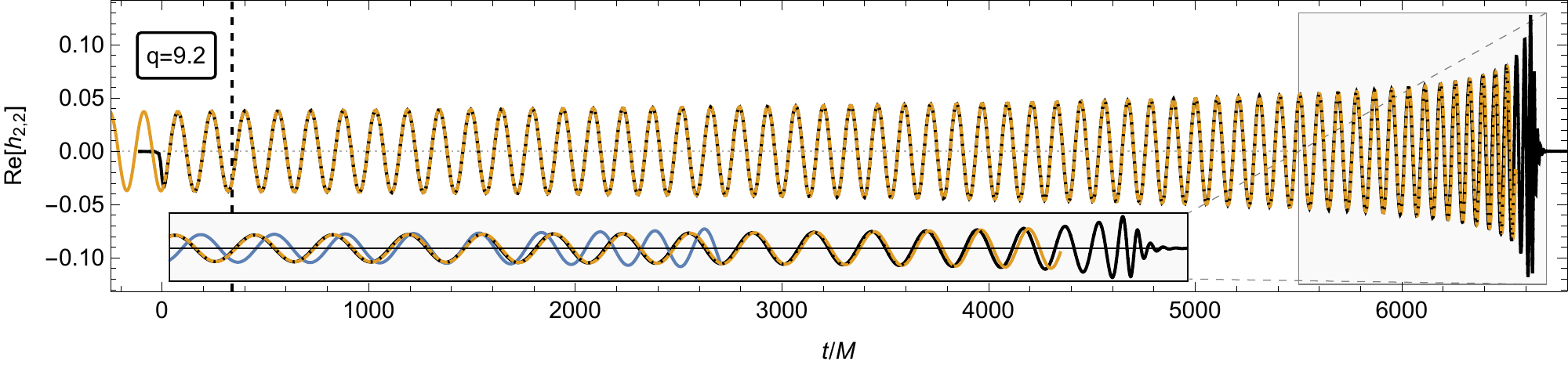}
	\includegraphics[width=0.95\textwidth]{waveform-q10-1PA.pdf}
	\caption{(Top to bottom) 1PAT1 (orange) and NR (black) waveforms for mass ratio 1:1, 1:4, 1:6, 1:9.2 and 1:10 nonspinning binaries. The inset shows a zoomed region near the merger and also shows the corresponding 0PA waveform (blue).}
\label{fig:NR-GSF}
\end{figure}
	
\newpage

\section{Comparisons between NR, GSF, PN and waveforms}
	
In this section we present a comparison between NR, GSF and PN waveforms.
For the PN waveform we use the TaylorT1 model \cite{Damour:2000zb}. The 1PAT1 GSF model significantly outperforms this PN waveform at $q=10$. Different PN approximants will have different levels of dephasing for any given mass ratio, but the dephasing in all approximants will grow roughly linearly with $q$ because of their non-negligible error in the 0PA term $F_0$ in Eq. (4).
	
\begin{figure}[!htb]
	\includegraphics[width=0.95\textwidth]{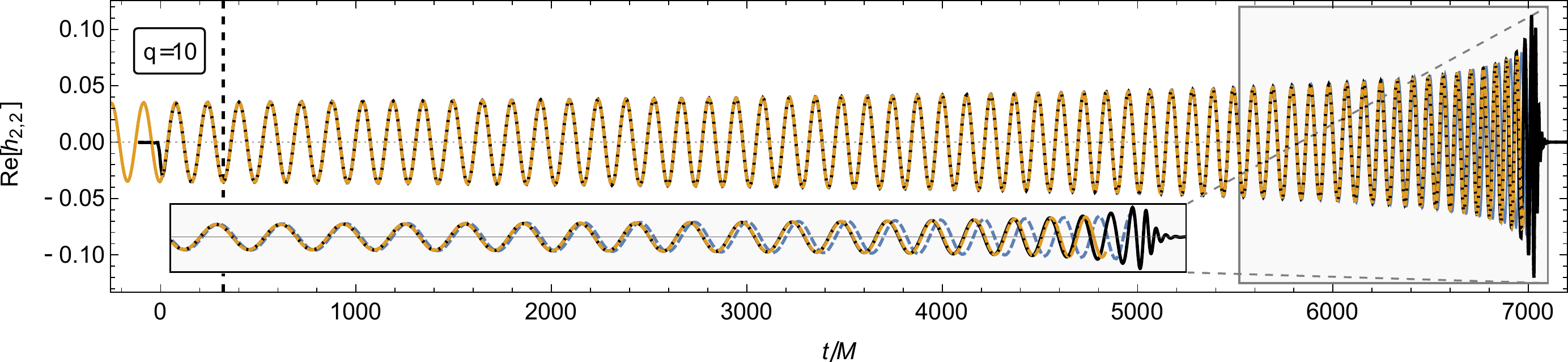}
	\caption{Comparison between waveforms computed by NR (black), 1PAT1 GSF (orange), and TaylorT1 PN (dashed blue) for a binary with $q=10$. The inset shows a zoomed region near the merger where it can be seen that the 1PAT1 GSF model has significantly less dephasing compared to the NR waveform than the TaylorT1 PN waveform.
	}\label{fig:NR-GSF-PN-waveforms}
\end{figure}	

\begin{figure}[!htb]
	\includegraphics[width=0.95\textwidth]{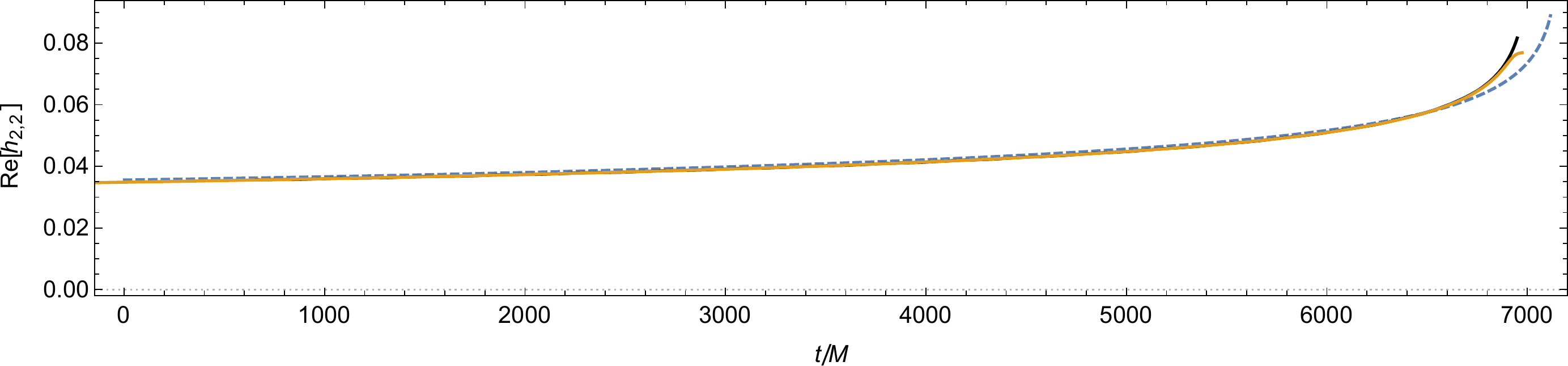}
	\caption{The same as Fig.~\ref{fig:NR-GSF-PN-waveforms} but comparing the amplitudes of the waveforms.}\label{fig:NR-GSF-PN-amplitude}
\end{figure}

\begin{figure}[!htb]
	\includegraphics[width=0.95\textwidth]{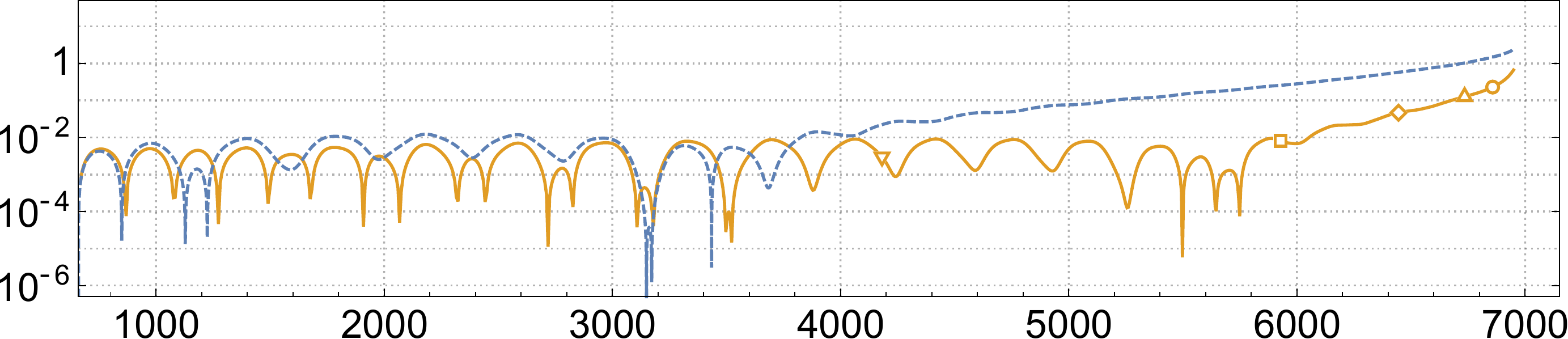}
	\caption{The same as Fig.~\ref{fig:NR-GSF-PN-waveforms} but showing the phase difference between the NR and 1PAT1 GSF waveforms (orange) and the phase difference between NR and TaylorT1 PN waveforms (dashed blue). The markers on the orange curve indicate the point on the waveform 32 (upside-down triangles), 16 (squares), 10 (diamonds), 6 (triangles), and 4 (circles) cycles before merger.}\label{fig:NR-GSF-PN-phase}
\end{figure}
		
\bibliography{bibfile}